\documentclass[twocolumn,10pt,cleanhead,cleanfoot]{asme2e}

\usepackage{xcolor}
\usepackage{graphicx}
\usepackage{amsmath} 
\usepackage{subcaption}
\usepackage{enumitem}
\usepackage{url}
\setlist[itemize]{label=\textbullet}
\confshortname{IDETC/CIE 2026}
\conffullname{the ASME 2026 International Design Engineering Technical Conferences \&\\
              Computers and Information in Engineering Conference}

\confdate{August 23-August 26}
\confyear{2026}
\confcity{Houston}
\confcountry{USA}

\papernum{IDETC/CIE 2026-193993}

\title{LLM4CAD-Editor: An Intent-Aware Large Language Model Framework for Multi-Level Computer-Aided Design Editing}

\author{
Yuewan Sun
    \affiliation{
	Walker Department of Mechanical Engineering\\
	University of Texas at Austin\\
	Austin, Texas 78712\\
    Email: sunyuewan2022@utexas.edu 
    }	
}

\author{Zhenghui Sha 
    \affiliation{Walker Department of Mechanical Engineering\\
	University of Texas at Austin\\
	Austin, Texas 78712\\
	Email: zsha@austin.utexas.edu
    }
}

\begin{document}

\maketitle    

\begin{abstract}
Large language models (LLMs) have recently enabled automatic generation of parametric computer-aided design (CAD) programs from natural language. However, real-world CAD workflows are inherently iterative and require reliable editing rather than one-shot model synthesis. In this work, we propose LLM4CAD-Editor, an LLM-based intent-aware framework for instruction-guided CAD editing based on a structured domain-specific language (LLM4CAD-DSL). The symbolic representation of LLM4CAD-DSL enables robust geometric modification through a feature-level entity selection mechanism, allowing models to reference geometry via feature names instead of coordinates, thus transforming fragile coordinate-based reasoning into natural language-based reasoning that many LLMs can handle. 
We construct a multimodal CAD editing dataset with over 35,139 instruction–program pairs via DSL-based augmentation and vision–language instruction synthesis, covering functional-, operation-, and parameter-level editing intents. To validate the work, we fine-tuned a 32B-parameter language model for DSL editing generation. Experimental results show high parsing accuracy for parameter-level edits (96.3\%) and strong intent satisfaction rates of 82\% for functional instructions. The model also achieves an average Intersection-over-Union (IoU) of 0.935 for parameter-level edits, 0.871 for operation-level edits, and 0.708 for functional-level edits, while the corresponding average editing distances are 0.176, 0.579, and 2.859, respectively. Comparative studies further demonstrate a significant improvement in editing robustness by 1.4x over Python-based CAD scripting approaches. These results confirm that LLM4CAD-Editor can reliably perform both low-level parameter modifications and high-level functional edits, maintaining high accuracy and low structural errors across diverse editing tasks.
\end{abstract}


\section{Introduction}
As Large Language Models (LLMs) continue to advance, they have been increasingly applied to computer-aided design (CAD) generation tasks. Recent approaches demonstrate that LLMs can translate natural language descriptions into parametric modeling sequences \cite{10.1115/1.4067085,wu2024cadvlmbridginglanguagevision,ZHOU2026104006}. However, despite promising progress, generated CAD models often lack geometric precision, parametric consistency, and fine-grained controllability. In this context, parametric consistency means maintaining valid relationships among parameters, features, and construction history, while fine-grained controllability means allowing localized edits to specific dimensions or operations without unintended global changes. These limitations have hindered the use of such models in practical engineering applications and have made them function more as concept generation tools than as reliable design systems. Therefore, there is a growing need for LLM-based CAD systems that support reliable, controllable editing after initial generation.

In practice, CAD model creation is inherently iterative. Designers typically begin with a coarse draft and progressively refine it through multiple rounds of modification until the model satisfies functional and manufacturing requirements. This underscores the increasing significance of CAD editing as a core capability. 
Compared to one-shot CAD generation, CAD editing better reflects real-world workflows, where adjustments, corrections, and refinements dominate the design cycle. Nevertheless, CAD editing remains underexplored in relation to CAD generation.

\begin{figure*}[t]
    \centering
    \includegraphics[width=0.8\textwidth]{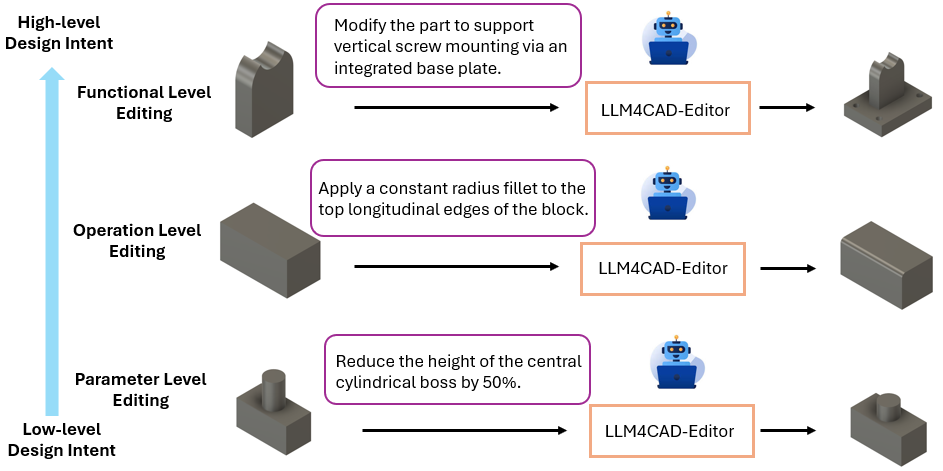}
    \caption{Overview of the LLM4CAD-Editor framework. The system takes natural language editing instructions at multiple abstraction levels (functional, operation, and parameter) and generates structured CAD edits using LLM4CAD-DSL.}
    \label{fig:overview}
\end{figure*}

Existing studies \cite{liu2025breplerlanguageguidededitingcad,yuan2025cadeditorlocatetheninfillframeworkautomated} on CAD editing face two major limitations. First, most current methods only support relatively simple geometric compositions derived from sketch–extrude sequences. Advanced parametric operations—such as revolve, fillet, chamfer, and cross-feature dependency updates—are rarely addressed. As a result, these systems struggle to handle industry-level design scenarios that require a deeper understanding of the modeling history and parametric relationships.

Second, natural language editing instructions play a crucial role throughout the editing process, serving as the primary medium of communication between humans and AI systems. However, users at different levels of expertise express design intent in fundamentally different ways, as shown in Figure~\ref{fig:overview}. Non-expert users often describe modifications at the functional level using abstract instructions, such as “make this part stronger” or “adjust it so it looks more balanced.” In contrast, professional designers tend to provide precise, operation-level commands, such as “increase the extrusion depth by 5mm” or “add a fillet with a 2mm radius to the outer edge.” Current studies largely overlook this distinction in design intent representation and fail to model the variability in linguistic abstraction between user groups.

This gap raises a key research question: Can LLMs support CAD editing that accommodates multi-level design intent—from high-level functional descriptions to low-level parametric operations—while supporting advanced modeling operations beyond simple sketch–extrude workflows? Addressing this challenge has the potential to bring LLM4CAD systems closer to practical deployment, enabling more flexible, intelligent, and user-adaptive design assistance.

In this work, we propose LLM4CAD-Editor, a unified framework for natural language-based CAD editing. Building upon LLM4CAD-DSL, the domain-specific language (DSL) representation developed in our previous work \cite{sun2026_llm4cad_dsl}, we represent CAD models as structured parametric programs with explicit feature-level CAD operation sequences (i.e., modeling histories) to enable precise and consistent modifications. 

To construct training data, we develop a data augmentation pipeline that converts LLM4CAD-DSL scripts into paired original–edited samples through controlled transformations. This pipeline systematically applies controlled modifications to generate diverse editing trajectories, allowing the model to learn structured transformations over parametric programs.

To model multi-level design intent, we employ the vision–language model Qwen3-VL-32B \cite{Qwen3-VL} to generate editing instructions at three levels of abstraction: (1) functional-level instructions, which describe high-level design goals; (2) operation-level instructions, which specify CAD operations; and (3) parameter-level instructions, which define explicit numerical or geometric changes. Based on this multi-level instruction corpus, we fine-tune Qwen3-32B \cite{qwen3technicalreport} using Low-Rank Adaptation (LoRA)\cite{hu2021loralowrankadaptationlarge} to adapt the model for intent-aware CAD editing.

We evaluate LLM4CAD-Editor across different instruction levels using four complementary metrics. Parsing rate is first used to measure the syntactic validity of generated DSL programs. Intersection-over-Union (IoU) is then employed to quantify geometric consistency between the edited and target models. Editing distance further evaluates the correctness of modeling operations at the program level. Finally, an LLM-assisted evaluation protocol is adopted to assess semantic alignment between generated edits and high-level design intent. Together, these evaluations demonstrate the model’s ability to reliably interpret and execute instructions across varying degrees of abstraction and design intent.

To summarize, our main contributions include:
\begin{itemize}

\item We introduce LLM4CAD-Editor, an intent-aware framework for instruction-guided CAD editing that bridges high-level functional descriptions and low-level parametric operations within a unified modeling pipeline.

\item We construct a multimodal CAD editing dataset with multi-level editing instructions, generated via DSL-based program augmentation and vision–language instruction synthesis, covering functional-, operation-, and parameter-level design intents.

\item We demonstrate through comprehensive experiments that the proposed framework achieves high editing accuracy and strong geometric consistency across different intent levels. Complementary qualitative comparisons further highlight the advantages of symbolic editing representations over geometry-driven Python-based CAD scripting.

\end{itemize}

\section{Related Work}
\subsection{LLM-based CAD Generation and Editing}
Recent advances in Large Language Models (LLMs) have enabled automatic CAD generation from natural language or image instructions, where textual or visual input is translated into parametric modeling sequences. Several works have demonstrated the potential of this approach. For example, CAD-GPT~\cite{kapsalis2024cadgptharnessingnaturallanguage} maps 3D positions and sketch-plane rotations into a language feature space to achieve precise spatial localization, supporting both text-to-CAD and image-to-CAD tasks. CAD-MLLM~\cite{xu2025cadmllmunifyingmultimodalityconditionedcad} fine-tunes a pre-trained LLM to perform unified conditional generation from text, images, and point clouds, aligning multimodal features with CAD command sequences. Similarly, CAD-Llama~\cite{li2025cadllamaleveraginglargelanguage} uses a hierarchical annotation pipeline to convert sketch–extrude sequences into a structured parametric code format, providing hierarchical semantic descriptions for LLM-based generation.

While these methods achieve impressive results in one-shot CAD generation, they remain primarily focused on producing initial drafts rather than supporting iterative refinement. Moreover, most generation approaches are limited in controllability and fine-grained parametric consistency. To address this, CAD-Editor~\cite{yuan2025cadeditorlocatetheninfillframeworkautomated} explores using LLMs to modify numerical CAD sequences based on natural language instructions. However, it largely handles basic sketch-extrude operations and struggles with complex feature dependencies.

Parallel to sequence-based methods, B-repLer~\cite{liu2025breplerlanguageguidededitingcad} introduces a latent-space editing framework that operates directly on B-rep data and complex freeform surfaces, bypassing the need for explicit construction history. Yet, by decoupling the editing process from the modeling sequence, it inherently lacks the fine-grained parametric controllability required for precision engineering. This makes it difficult to maintain strict geometric constraints or synchronize logical dependencies between interdependent features. These limitations motivate the development of CAD editing frameworks that retain structured operation sequences and feature dependencies, allowing reliable geometric modification and effective support for advanced parametric operations beyond extrusion.

\subsection{Natural Language Interfaces and Multi-Level Design Intent}

Natural language serves as a pivotal interface in CAD editing, aiming to bridge the semantic gap between high-level design intent and low-level geometric operations. Recent exploratory studies have begun to leverage LLMs and multimodal architectures to interpret textual instructions for parametric modification. For instance, CAD-Editor~\cite{yuan2025cadeditorlocatetheninfillframeworkautomated} and B-repLer~\cite{liu2025breplerlanguageguidededitingcad} demonstrate the potential of translating natural language commands into localized geometric updates or numerical sequence adjustments. These approaches have shown preliminary success in basic functional tasks, such as resizing or translating elementary features, suggesting a path toward enabling users to perform design iterations without deep CAD expertise.

However, current methods typically treat all instructions uniformly, without considering differences in user expertise and user intent. In practice, non-expert users often issue high-level or functional instructions, such as “make this part stronger” or “adjust the design to be more balanced,” which do not correspond directly to specific parametric operations. Professional designers, in contrast, tend to provide precise, operation-level instructions with numerical or geometric specifications, such as “increase the extrusion depth by 5mm” or “add a fillet with a 2mm radius to the outer edge.” Existing CAD editing systems fail to model these differences in design intent, limiting their adaptability across user groups. Furthermore, there remains a significant challenge in aligning these multi-level instructions within CAD editing workflows, particularly in understanding how high-level functional intent relates to specific parametric adjustments. Addressing this misalignment is critical for developing more flexible CAD editing systems that can better accommodate the diverse communication styles of both novice and expert users.

Overall, prior CAD editing approaches either provide limited support for advanced parametric operations or do not explicitly capture multi-level design intent conveyed through natural language instructions. In the following section, we present LLM4CAD-Editor to address these limitations.

\section{Method}
\subsection{Methodology Overview}
The core of our CAD editing framework leverages the structural advantages of LLM4CAD-DSL (detailed in Section \ref{sec:llm4cad_dsl}). The methodology follows a three-stage pipeline: data synthesis, multi-level instruction generation, and supervised fine-tuning. 

First, we perform data augmentation on the original LLM4CAD-DSL dataset to construct a specialized sub-dataset for editing tasks. This process involves generating paired DSL scripts (\textit{original} and \textit{edited}) that represent valid geometric transformations (Section \ref{sec:dsl_pair}). Second, to bridge the gap between geometric changes and natural language intent, we utilize a Vision-Language Model (VLM), specifically Qwen3-VL-32B, to synthesize editing instructions across three distinct semantic levels: \textit{Parameter Level}, \textit{Operation Level}, and \textit{Functional Level}. This results in a comprehensive dataset of editing triplets: $\langle \text{editing instruction, original DSL, edited DSL} \rangle$ (Section \ref{sec:editing_dataset}). Finally, we fine-tune a Large Language Model (Qwen3-32B) on this triplet dataset to develop the LLM4CAD-Editor, enabling the model to accurately map natural language design intent to precise, stable DSL modifications.

\subsection{LLM4CAD-DSL: LLM-Friendly DSL for Robust CAD Editing}
\label{sec:llm4cad_dsl}

\begin{table*}[t]
\centering
\small
\setlength{\tabcolsep}{6pt}
\renewcommand{\arraystretch}{1.15}
\begin{tabular}{p{0.22\textwidth} p{0.74\textwidth}}
\hline
\textbf{Operation} & \textbf{Example} \\
\hline
Define sketch &
\texttt{sketch:sketch\_1(origin=(-0.23, -0.09, 0.00), normal=(0.0000, 0.0000, 1.0000))} \\

Define variable &
\texttt{set var edge\_2=edge(revolve\_1:ridge(vertex(sketch\_1:v\_2)))} \newline
\texttt{set var face\_1=face(extrude\_1:side(line(sketch\_1:arc\_1)))} \\

Add line &
\texttt{vertex:v\_1=(0.0000, 0.0000)} \newline
\texttt{vertex:v\_2=(0.7500, 0.0000)} \newline
\texttt{line:line\_1(start=v\_1, end=v\_2)} \\

Add arc &
\texttt{arc:arc\_2(start=v\_12, end=v\_13, center=v\_14)} \\

Add circle &
\texttt{circle:circle\_1(center=vertex\_1, radius=0.0103)} \\

Add loop &
\texttt{loop:loop\_2(entities=[line\_11, arc\_13, line\_14, circle\_16])} \\

Extrude &
\texttt{extrude:extrude\_1(entity=sketch\_1, length1=-0.75, length2=0.00)} \\

Revolve &
\texttt{revolve:revolve\_1(entity=sketch\_1, axis=line\_1, angle1=360.00, angle2=0.00)} \\

Fillet &
\texttt{fillet:fillet\_1(entities=[edge\_5, edge\_6, edge\_7], radius=0.04)} \\

Chamfer &
\texttt{chamfer:chamfer\_1(entities=[edge\_1, face\_2], distance1=0.01, distance2=0.02)} \\

Pocket &
\texttt{pocket:pocket\_1(entity=sketch\_2, length1=0.34, length2=0.00)} \\

Groove &
\texttt{groove:groove\_1(entity=sketch\_1, axis=line\_1, angle1=360.00, angle2=0.00)} \\
\hline
\end{tabular}
\caption{Examples of supported LLM4CAD-DSL operations and their corresponding command syntax.}
\label{tab:LLM4CAD-DSL_examples_full}
\end{table*}

The foundation of the editing task in this study is built upon our previous development of an LLM-friendly domain-specific language (DSL) for CAD: LLM4CAD-DSL. The architecture of LLM4CAD-DSL is specifically designed to bridge the gap between high-level semantic intent and low-level geometric execution, serving as a robust substrate for automated CAD editing. Traditional Python-based CAD APIs, such as CadQuery \cite{cadquery2026} and FreeCAD \cite{freecad}, often rely on absolute 3D coordinates. In contrast, LLM4CAD-DSL introduces a symbolic selection mechanism that assigns unique and interpretable semantic labels to geometric entities such as faces, edges, and loops. In an editing context, this enables the LLM to modify a model by referencing stable identifiers---such as \texttt{Edge,Topcap,Loop}---rather than calculating precise $(x, y, z)$ coordinates, a task where LLMs frequently suffer from geometric hallucinations. By decoupling feature definitions from rigid spatial constraints, the DSL ensures that the editing process remains focused on topological relationships, significantly reducing the reasoning complexity required for localized modifications. The syntax and representative examples of LLM4CAD-DSL are detailed in Table~\ref{tab:LLM4CAD-DSL_examples_full}.

Furthermore, the structured and modular nature of LLM4CAD-DSL facilitates topology-aware updates that are essential for iterative design. In traditional programmatic CAD, minor parameter changes often trigger the ``Topological Naming Problem,'' where downstream operations fail because their geometric references have shifted or vanished. LLM4CAD-DSL mitigates this fragility by maintaining consistent symbolic references throughout the construction sequence. This independence allows the LLM4CAD-Editor to perform ``surgical'' edits---such as adjusting a primitive parameter or inserting a complex functional feature like a mounting plate---without triggering a cascade of failures in subsequent operations. Consequently, the DSL provides a unified representation that supports a hierarchy of editing intents, ranging from low-level geometric tuning to high-level functional additions, ensuring both the reliability and the interpretability of the edited CAD models.

To leverage these structural advantages in a data-driven editing framework, we employ the LLM4CAD-DSL dataset \cite{sun2026_llm4cad_dataset} 
as the primary source and benchmark for all editing tasks in this work. This large-scale dataset comprises 107,994 high-fidelity CAD models, each meticulously converted from the WHUCAD dataset \cite{WOS:001360811000005} into their corresponding LLM4CAD-DSL representations. Unlike existing datasets that are often limited to simple geometric primitives, the LLM4CAD-DSL dataset encompasses a wide range of advanced industrial features, including \textit{revolve, fillet, chamfer, groove, and pocket} operations. Each entry in the dataset provides aligned DSL scripts, valid STL models generated from CAD programs with manufacturing-relevant operations, and multi-view rendered images, offering a rich multimodal context for the LLM4CAD-Editor. By capturing complex construction sequences and their resulting topological structures, this dataset provides a diverse and representative sample of realistic engineering designs, which is essential for training the model to recognize and manipulate sophisticated geometric features during the iterative editing process.

\subsection{Data Synthesis}
\label{sec:data_construction}
\subsubsection{DSL Augmentation}
\label{sec:dsl_pair}

\begin{figure}[htbp]
    \centering
    
    \begin{subfigure}{\columnwidth}
        \centering
        \includegraphics[width=\linewidth]{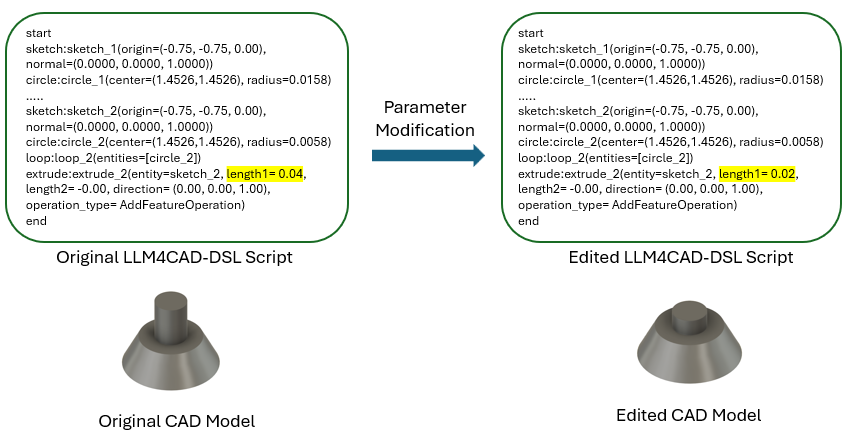}
        \caption{Example of parameter modification.}
        \label{fig:change_parameter}
    \end{subfigure}
    
    \vspace{0.3cm}
    
    \begin{subfigure}{\columnwidth}
        \centering
        \includegraphics[width=\linewidth]{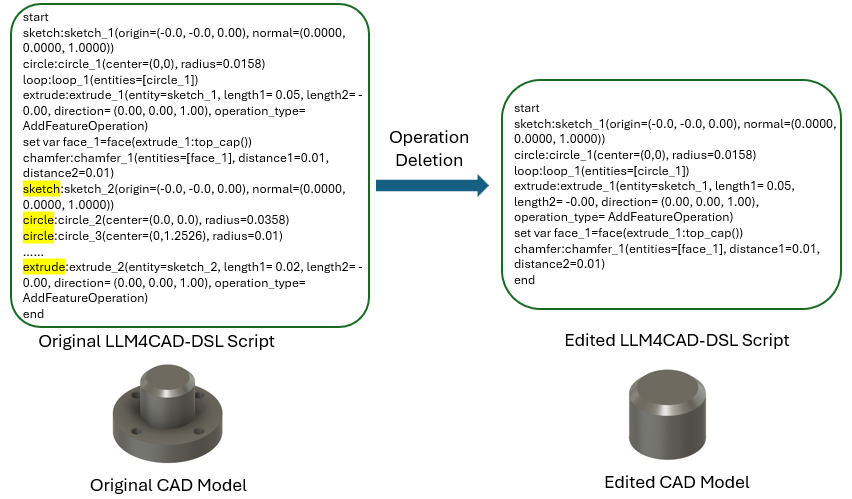}
        \caption{Example of operation deletion.}
        \label{fig:change_operation}
    \end{subfigure}
    
    \caption{Two types of data augmentation applied to the LLM4CAD-DSL dataset.}
    \label{fig:ab}
\end{figure}

Based on the LLM4CAD-DSL dataset, we design two types of data augmentation strategies to construct CAD editing pairs.

\textbf{1) Parameter Modification.}
In the first strategy, we randomly modify numerical parameters within DSL operations. These parameters include, for example, circle radius, extrusion length, fillet radius, chamfer dimensions, as well as the selected edges for fillet and chamfer operations. As illustrated in Figure~\ref{fig:change_parameter}, the highlighted extrusion length is reduced to half of its original value. In general, each numerical parameter is randomly scaled within a range of 0.5$\times$ to 2$\times$ its original value. This approach preserves the structural validity of the DSL program while generating meaningful geometric variations.

\textbf{2) Operation Deletion.}
The second strategy focuses on structural edits. We randomly select one line of DSL code for deletion. Based on the hierarchical structure of the DSL and geometry dependencies, we automatically detect all subsequent operations that depend on the deleted geometry (e.g., via referenced geometry names). These dependent operations are removed accordingly to maintain the validity and executability of the CAD model. An example is shown in Figure~\ref{fig:change_operation}.

It should be noted that each original–edited DSL pair is interchangeable. During the subsequent generation of the editing instruction data for training, we swap the source and target programs. Therefore, the learning task includes not only deletion-based editing but also addition-based editing.

\begin{figure}[h]
    \centering
    \includegraphics[width=\columnwidth]{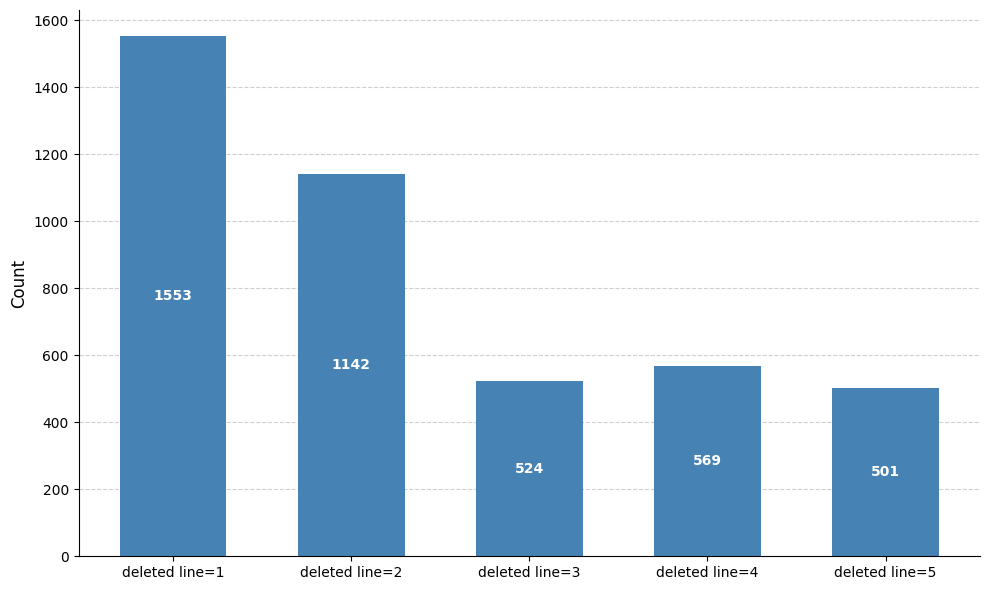}
    \caption{The distribution of deleted line counts.}
    \label{fig:deleted_line_distribution}
\end{figure}

For each augmented pair, we record the number of deleted lines, which corresponds to the edit distance between the original and edited DSL programs. This edit distance serves as a proxy for the complexity of the editing task. After augmentation, all edited DSL programs are rendered to further verify their validity. This validation ensures that the augmented programs remain executable and produce valid CAD geometries. Because the source DSL programs contain manufacturing-relevant operations such as revolve, fillet, chamfer, groove, and pocket, the augmented dataset also preserves engineering-oriented geometric features that remain suitable for manufacturing-related CAD modeling tasks. Through parameter modification, we obtain 8,973 DSL pairs. Through operation deletion, we obtain 4,289 DSL pairs. The distribution of deleted line counts is shown in Figure~\ref{fig:deleted_line_distribution}.

\subsubsection{Instruction Generation}
\label{sec:instruction}

\begin{figure}[h]
\centering
\includegraphics[width=\columnwidth]{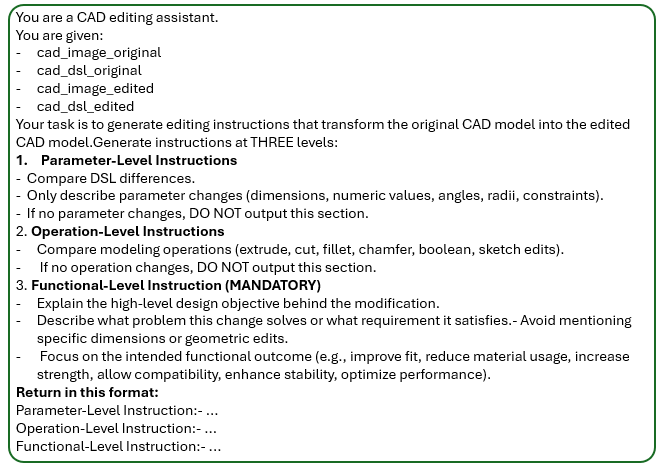}
\caption{System prompt used for instruction generation.}
\label{fig:instruction_system_prompt}
\end{figure}

\begin{figure*}[h]
\centering
\includegraphics[width=0.75\textwidth]{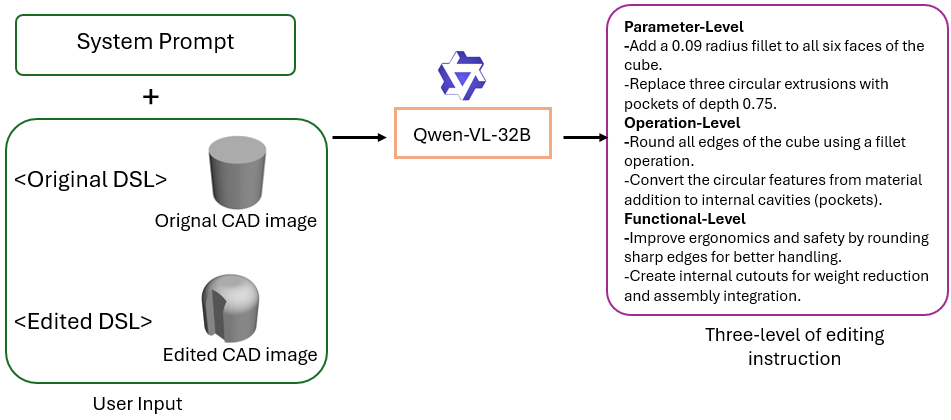}
\caption{Pipeline for editing instruction generation.}
\label{fig:instruction_generation}
\end{figure*}

To generate editing instructions for each DSL pair, we employ Qwen3-VL-32B as the instruction generation model. For each DSL program, we first render 8 isometric views of the corresponding CAD model. We then compute the Structural Similarity Index Measure (SSIM) \cite{larkin2015structuralsimilarityindexssimplified} between the rendered images of the original and edited models and select the viewpoint with the largest visual difference. This ensures that the geometric changes introduced by the edit are clearly captured in the selected image.

Next, we provide the vision-language model with a system prompt (Figure \ref{fig:instruction_system_prompt}) together with the original DSL, the rendered image of the original CAD model, the edited DSL, and the rendered image of the edited CAD model. Based on this multimodal input, the model generates an editing instruction describing the transformation from the original model to the edited model. The overall instruction generation pipeline is illustrated in Figure~\ref{fig:instruction_generation}.

\begin{figure}[h]
    \centering
    \includegraphics[width=\columnwidth]{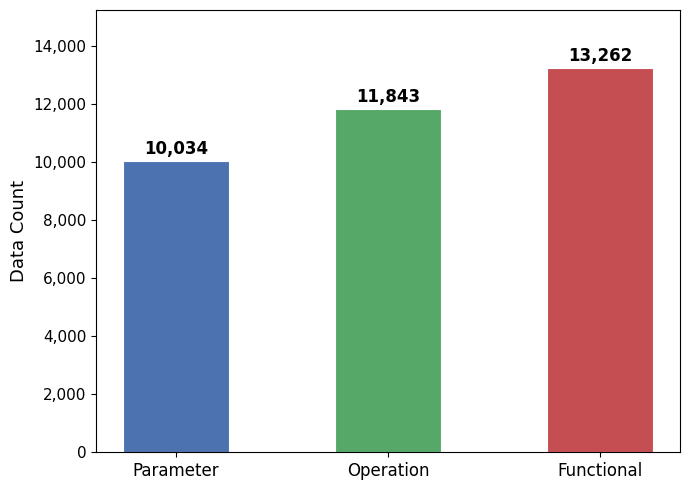}
    \caption{Distribution of the editing dataset across three levels of design intent.}
    \label{fig:editing_dataset_distribution}
\end{figure}

The generated instruction is structured into three levels of intent: functional-level, parameter-level, and operation-level descriptions. The functional-level instruction is mandatory and captures the high-level design intent, while the parameter-level and operation-level descriptions are optionally generated by the model depending on the specific edit. In total, the dataset contains 13,262 editing pairs and 35,139 instructions. 
The distribution of instructions across the three levels is shown in Figure~\ref{fig:editing_dataset_distribution}.

\subsubsection{Editing Dataset}
\label{sec:editing_dataset}

\begin{table*}[p]
    \centering
    \includegraphics[width=0.7\textwidth]{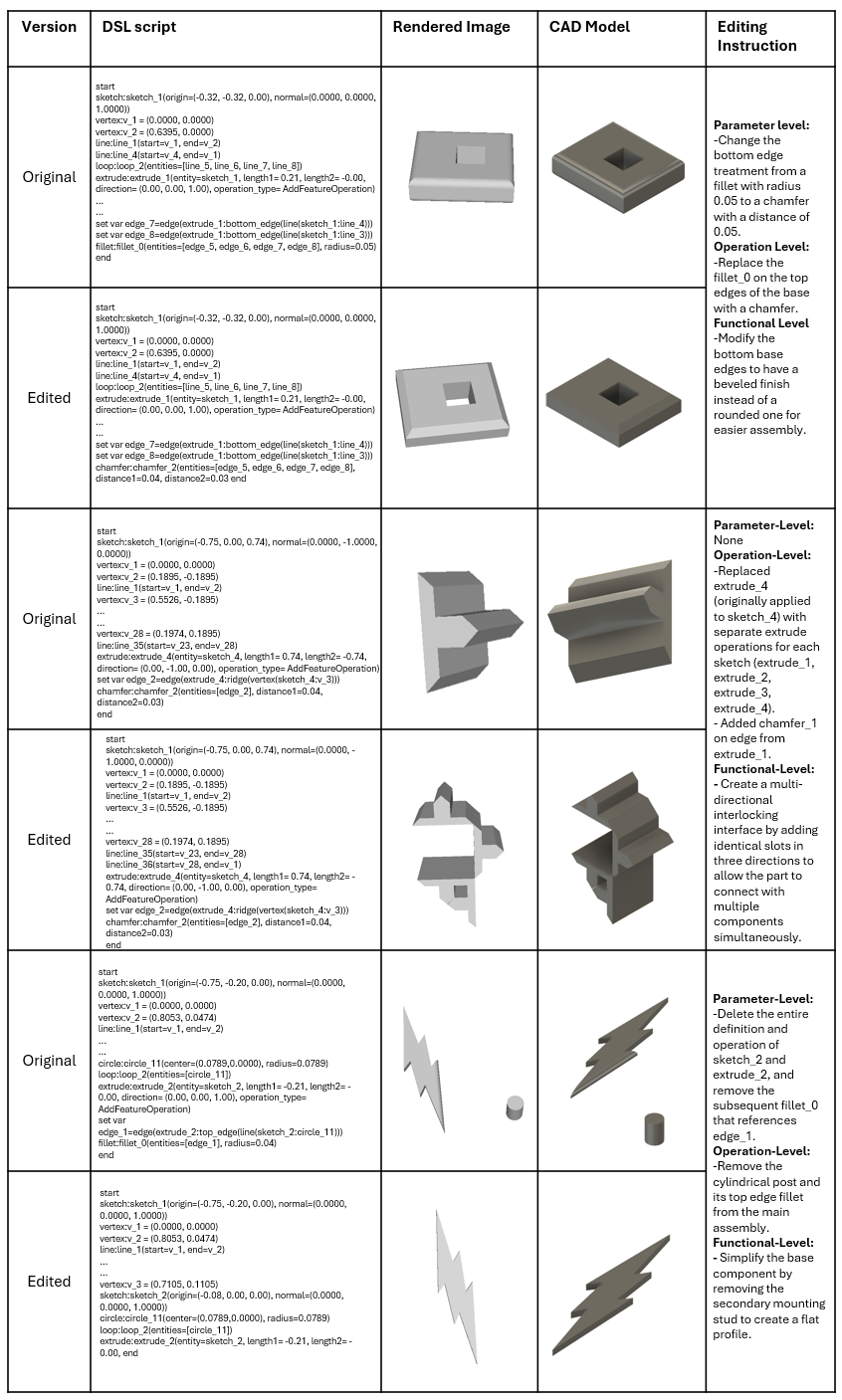}
    \caption{Overview of the editing dataset.}
    \label{tab:dataset_overview}
\end{table*}

After generating editing instructions with three levels of design intent, we construct the final CAD editing dataset. 
Each dataset entry corresponds to an editing pair derived from the augmented DSL scripts together with the generated editing instructions. 
As illustrated in Table~\ref{tab:dataset_overview}, each editing pair consists of an editing instruction and two CAD data points: an \textit{original} data point and an \textit{edited} data point. 
More specifically, each pair consists of the following elements:

\begin{itemize}
\item an editing instruction, structured into three levels of design, 
\item the original DSL script,
\item the rendered image of the original CAD model,
\item the original CAD model,
\item the edited DSL script,
\item the rendered image of the edited CAD model,
\item the edited CAD model.
\end{itemize}


Importantly, the dataset is inherently multimodal, as each editing pair contains DSL scripts, rendered images, and CAD models. 
This design makes the dataset suitable not only for text-based instruction-following, but also for future research on multimodal CAD editing, such as image-conditioned editing, vision-language instruction grounding, and multimodal CAD generation.

\section{Experiment}
\subsection{Experimental Setup}

\begin{figure*}[h]
    \centering
    \includegraphics[width=0.8\textwidth]{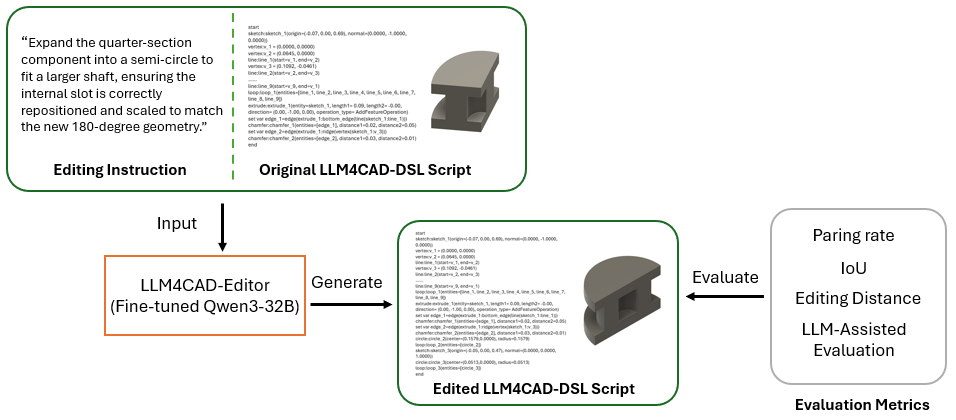}
    \caption{Evaluation process of LLM4CAD-Editor.}
    \label{fig:experiment_workflow}
\end{figure*}

Based on the editing dataset constructed in Section~\ref{sec:editing_dataset}, we fine-tune Qwen3-32B for instruction-guided CAD editing. During training, each editing instruction is treated as an independent instance. 
Although the constructed dataset is inherently multimodal and includes rendered images and 3D CAD models, the current study focuses on a purely text-based editing setting. Each training sample consists of a triplet: an editing instruction (at a specific intent level), the original DSL script, and the target edited DSL script. Visual information is not used during training or inference. This design allows us to isolate and evaluate the capability of large language models to perform structured CAD editing through symbolic program reasoning alone, without relying on visual grounding signals. The resulting dataset contains 32,139 training samples and 3,000 testing samples.

We fine-tune Qwen3-32B using LoRA \cite{hu2021loralowrankadaptationlarge} adapters applied to the projection and feed-forward layers (rank 16, $\alpha$ = 32, dropout = 0.05), with gradient checkpointing enabled to reduce memory consumption. Training is conducted on eight NVIDIA A100-80GB GPUs for 19.7 hours, using a maximum sequence length of 2048, a per-device batch size of 2, and a cosine learning rate schedule with a peak learning rate of $1\times10^{-4}$ over three epochs. The final model and tokenizer are saved for downstream evaluation.

\subsection{Metrics}

During inference, the fine-tuned LLM takes an editing instruction and the original DSL script as input, and generates the edited DSL script as output. An overview of this process is illustrated in Figure~\ref{fig:experiment_workflow}. The CAD renderings shown in the figure are provided for illustration purposes only and are not used by the model during inference. We evaluate the performance of the model using the following metrics.

\textbf{Parsing Rate.} Parsing rate measures the proportion of generated DSL scripts that can be successfully parsed into executable CAD models. Since LLM4CAD-DSL is a newly introduced DSL that the base model has not been exposed to during pretraining, this metric reflects whether the fine-tuned model has learned the syntax and structural rules of the DSL. A high parsing rate is also important for practical usability, as only syntactically valid scripts can be executed to produce CAD models.

\textbf{Intersection-over-Union (IoU).} For successfully parsed scripts, we compute the Intersection-over-Union (IoU) between the generated CAD model and the ground-truth CAD model to evaluate geometric accuracy. However, for high-level functional instructions that do not explicitly specify parameters, multiple valid designs may satisfy the same instruction. In such cases, IoU becomes a strict metric and may underestimate the quality of the generated edits.

\textbf{Editing Distance.} Editing distance measures the difference between the generated operation sequence and the ground-truth operation sequence in the DSL scripts. For instructions that do not involve numerical parameters, correctness of the operation sequence largely determines whether the generated CAD model aligns with the intended target design. When the operation sequence is correct, users can easily adjust numerical parameters in the DSL script with minimal effort. In this work, editing distance is computed using a dynamic programming (DP) algorithm \cite{10.1145/375360.375365} that finds the minimum number of insertions, deletions, and substitutions required to transform the generated sequence into the ground-truth sequence.

\begin{figure}[h]
    \centering
    \includegraphics[width=\columnwidth]{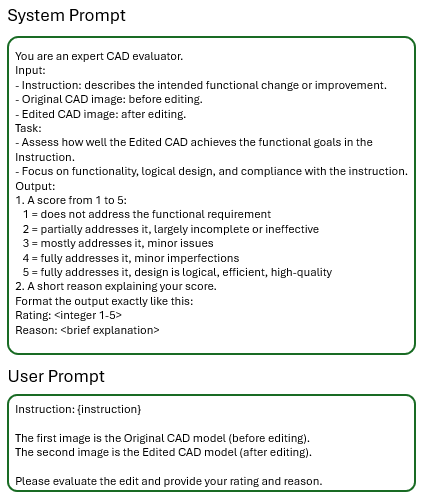}
    \caption{The system prompt and user prompt for LLM-Assisted Evaluation.}
    \label{fig:llm_eval_prompt}
\end{figure}

\textbf{LLM-Assisted Evaluation.} Because high-level functional instructions may admit multiple valid solutions, automatic metrics such as IoU cannot fully capture instruction adherence. To address this limitation, we adopt an LLM-assisted evaluation strategy for functional-level edits. Specifically, we use Llama-3.2-11B-Vision-Instruct \cite{grattafiori2024llama3herd} as an evaluator that receives the editing instruction, the rendered image of the original CAD model, and the rendered image of the edited CAD model. The evaluator rates the edit on a scale from 1 (worst) to 5 (best) based on how well the edited design satisfies the functional requirements described in the instruction, with emphasis on functionality, logical design, and compliance with the instruction. In addition to the score, the evaluator also provides a brief explanation for the rating. The system and user prompts used for this evaluation are shown in Figure~\ref{fig:llm_eval_prompt}.

\section{Results}
\subsection{Parsing Rate Results}

\begin{figure}[h]
    \centering
    \includegraphics[width=\columnwidth]{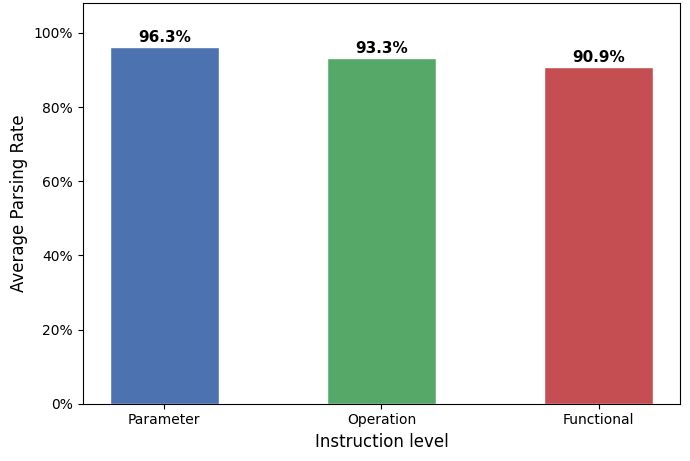}
    \caption{The average parsing rate for each instruction level.}
    \label{fig:parsing_rate_results}
\end{figure}

Figure~\ref{fig:parsing_rate_results} presents the average parsing rate for each instruction level. Among the three levels, parameter-level instructions achieve the highest parsing rate at 96.3\%, followed by operation-level instructions at 93.3\%. Functional-level instructions obtain a parsing rate of 90.9\%, which is slightly lower but still remains above 90\%.

The consistently high parsing rates across all instruction levels indicate that the fine-tuned model has successfully learned the syntax and structural conventions of LLM4CAD-DSL, including the naming rules used for selecting geometric entities. Notably, during inference we do not provide any in-context examples demonstrating the DSL syntax. The model must directly modify the original script and generate valid DSL programs on its own.

These results demonstrate that the model can reliably produce syntactically valid CAD DSL scripts, which is essential for practical deployment. A high parsing rate ensures that the generated outputs can be executed by the downstream CAD pipeline without requiring additional manual correction.

\subsection{IoU Results}

\begin{figure}[h]
\centering
\includegraphics[width=\columnwidth]{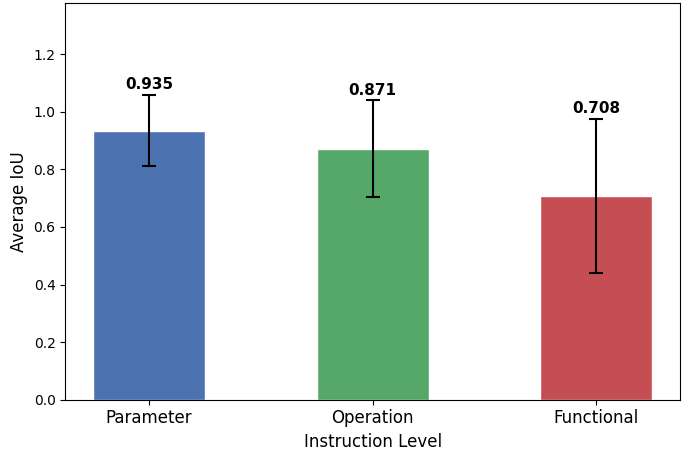}
\caption{The average IoU results for each instruction level.}
\label{fig:IoU_results}
\end{figure}

Figure~\ref{fig:IoU_results} shows the IoU results for three levels of editing instructions. The parameter-level instructions achieve the highest IoU score of 0.935, followed by the operation-level instructions with 0.871, while the functional-level instructions reach 0.708. 

The high IoU score for the parameter level indicates that the model can accurately edit the CAD DSL script when detailed instructions are provided. The performance of the operation level is lower because, although some instructions mention the operation name, they do not provide detailed numerical parameters for the desired CAD model. This ambiguity leads to a decrease in the IoU score.

The decreasing IoU across the three levels suggests that increased ambiguity in the instructions negatively affects the editing accuracy. However, it should be noted that for high-level instructions, the IoU metric may be overly strict. Overall, these results indicate that the model performs reliably when precise editing instructions are provided, while higher-level and more abstract instructions introduce ambiguity that makes accurate geometric editing more challenging. Statistical analysis shows that the parameter-level IoU is significantly higher than the functional-level score (p=0.014), and the operation-level IoU is also significantly higher than the functional-level score (p=0.025, paired t-test).

\subsection{Editing Distance Results}

\begin{figure}[h]
    \centering
    \includegraphics[width=\columnwidth]{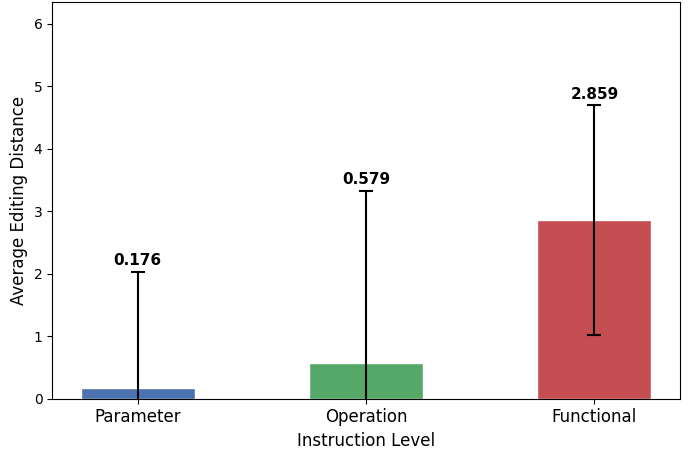}
    \caption{The editing distance for each instruction level.}
    \label{fig:editing_distance_level_results}
\end{figure}

Figure~\ref{fig:editing_distance_level_results} presents the average editing distance for each instruction level. 
Parameter-level instructions achieve the lowest editing distance of 0.176, followed by operation-level instructions with 0.579, while functional-level instructions have a substantially higher editing distance of 2.859. 

The low editing distances for parameter-level and operation-level instructions indicate that the generated DSL scripts are already very close to the ground-truth edits. In many cases, the editing distance is below 1, meaning that users would need to modify fewer than one line of code on average to obtain the correct script. This suggests that the model can accurately reproduce the intended operation sequence and parameters for low-level editing tasks. Statistical analysis shows that the editing distances for both parameter-level and operation-level instructions are significantly smaller than those for functional-level instructions (p \textless 0.05, paired t-test).

In contrast, functional-level instructions exhibit a significantly larger editing distance. This is expected because functional instructions express high-level and abstract design intents, which may correspond to multiple valid operation sequences. As a result, evaluating such edits solely based on the exact operation sequence becomes more challenging and may underestimate the quality of the generated designs.

\begin{figure}[h]
    \centering
    \includegraphics[width=\columnwidth]{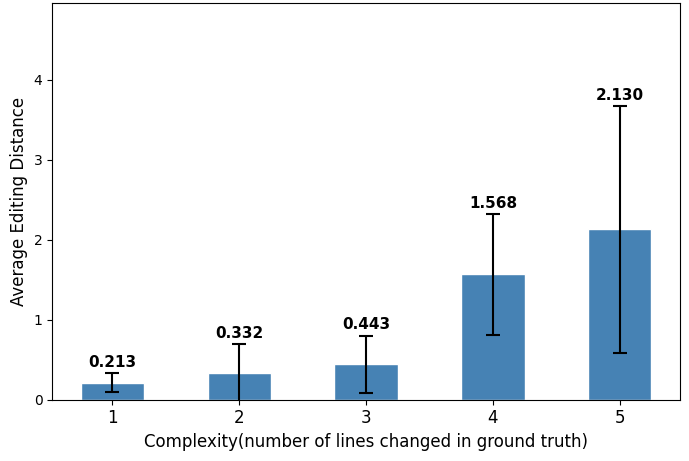}
    \caption{The editing distance by complexity level.}
    \label{fig:editing_distance_complexity_results}
\end{figure}

Figure~\ref{fig:editing_distance_complexity_results} further analyzes editing distance with respect to task complexity. In this study, we measure editing complexity by the number of lines changed between the original and ground-truth DSL scripts. As shown in the figure, the average editing distance increases as the editing complexity grows. This trend indicates that tasks requiring more extensive modifications are inherently more difficult for the model to reproduce exactly.

Overall, these results demonstrate that the model performs very well on low-level editing tasks, where the intended modifications are explicit and localized. In these cases, editing distance provides a reliable measure of correctness, and the low values indicate that the generated DSL scripts closely match the ground-truth edits. For high-level functional instructions, however, the intended design change is more abstract and may correspond to multiple valid operation sequences. As a result, evaluating such edits solely based on editing distance may not fully reflect their quality. To better assess the effectiveness of functional-level edits, we further conduct an LLM-assisted evaluation.

\subsection{LLM-Assisted Evaluation Results}

\begin{figure}[h]
    \centering
    \includegraphics[width=\columnwidth]{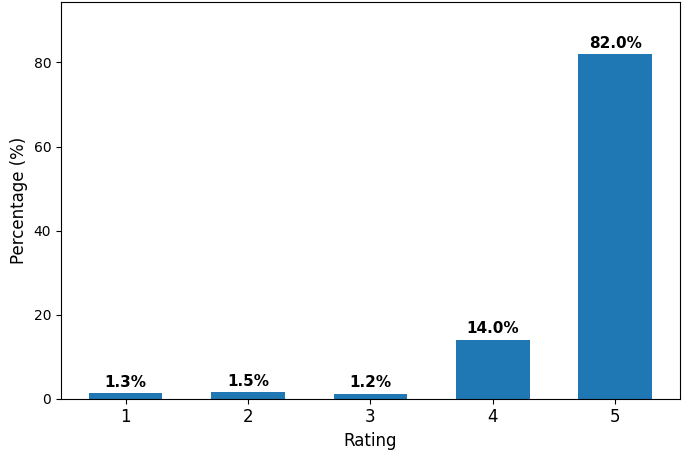}
    \caption{The rating distribution for LLM-assisted evaluation.}
    \label{fig:llm_eval_results}
\end{figure}

Since high-level functional instructions cannot be fully evaluated using automatic metrics, we adopt an LLM-assisted evaluation method to assess the quality of functional-level editing results. The rating scale ranges from 1 to 5, where 1 indicates that the generated edit fails to address the functional requirement, and 5 indicates that the edited design fully satisfies the instruction with a logical and high-quality solution. Ratings of 4 and 5 correspond to edits that successfully achieve the intended functional change, with only minor imperfections in some cases.

Figure~\ref{fig:llm_eval_results} shows the rating distribution of the generated edits. The majority of edits (82.0\%) receive the highest rating of 5, while 14.0\% receive a rating of 4. Only 4\% of the results obtain a rating below 4. Overall, the average rating across all functional-level edits is 4.375, reflecting consistently high-quality outputs.

These results indicate that the fine-tuned LLM is able to effectively follow high-level functional instructions and generate designs that satisfy the intended functional goals in most cases. Combined with the strong performance on low-level editing tasks reported in the previous sections, the model demonstrates promising capability for instruction-guided CAD editing across different levels of design intent. For examples of qualitative results, see Appendix~\ref{app:qualitative_assessment}.

\section{Discussion}
\subsection{Intent-Aware CAD Editing}

The experimental results highlight the importance of design intent representation in instruction-guided CAD editing. Our study organizes editing instructions into three levels of design intent—parameter-level, operation-level, and functional-level—which correspond to different levels of abstraction in design reasoning. The results show a clear relationship between the granularity of design intent and the difficulty of the editing task.

For parameter-level instructions, the editing intent is explicit and localized. These tasks typically involve modifying numerical parameters associated with existing geometric operations, such as extrusion height or hole radius. Because the required modification is clearly specified, the model achieves the highest performance across parsing rate, IoU, and editing distance.

For operation-level instructions, the model must identify the appropriate location and context in which to apply a specified geometric operation. Although the operation type is provided in the instruction, the model still needs to reason about entity selection and procedural dependencies within the CAD program. As a result, these tasks require a deeper understanding of the modeling sequence compared to parameter-level edits.

In contrast, functional-level instructions express high-level design goals rather than explicit modeling operations. To satisfy such instructions, the model must infer a sequence of geometric modifications that achieves the intended functional objective. Because multiple valid editing strategies may exist, traditional automatic metrics may underestimate the quality of the generated designs. The strong results from LLM-assisted evaluation indicate that the model can successfully translate high-level design intent into plausible CAD modifications in most cases.

This trend is consistent with broader findings in applied LLM research, where ambiguous or underspecified user instructions have been shown to reduce model reliability and increase the need for clarification or additional context. For example, CLAMBER reports that current LLMs have limited practical ability to identify and clarify ambiguous user queries~\cite{zhang2024clamber}, while AmbigNLG shows that instruction ambiguity can degrade generation quality and that resolving ambiguity improves alignment with user expectations~\cite{niwa2024ambignlg}. In the CAD domain, CADialogue similarly observes that semantically complex prompts benefit from additional visual input and human-in-the-loop refinement~\cite{ZHOU2026104006}. Therefore, the lower automatic scores observed for functional-level CAD editing should be interpreted as part of a broader challenge in human-LLM interaction, rather than as a limitation unique to the specific model used in this work.

These findings suggest a paradigm of intent-aware CAD editing, where CAD systems interpret user instructions at different levels of design abstraction and generate appropriate modeling operations accordingly. Such a framework can enable more natural interaction between users and CAD systems, allowing both expert designers and non-expert users to modify complex models using high-level design intents rather than low-level geometric commands.

\subsection{Comparison of CAD Representations}

\begin{figure}[h]
    \centering
    \includegraphics[width=\columnwidth]{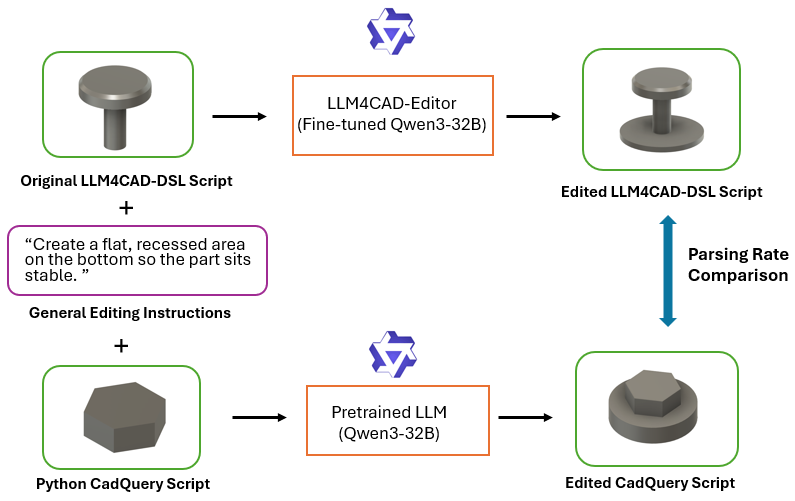}
    \caption{Experimental setup for comparing editing support between LLM4CAD-DSL and the Python-based CadQuery representation.}
    \label{fig:comparison_experiment}
\end{figure}

\begin{figure}[h]
    \centering
    \includegraphics[width=0.8\columnwidth]{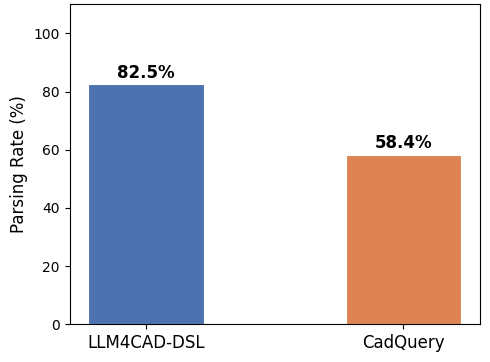}
    \caption{Parsing rate comparison between LLM4CAD-DSL and CadQuery under instruction-guided editing.}
    \label{fig:comparison_parsing_rate}
\end{figure}



To further evaluate the advantages of LLM4CAD-DSL for editing tasks, we compare it with a commonly used Python-based CAD representation in the literature, CadQuery \cite{cadquery2026}. 

A key structural difference between the two representations lies in how geometric entities are referenced and preserved during editing. In LLM4CAD-DSL, downstream features can remain valid even when parameters of earlier operations are modified due to the feature-level entity selection mechanism. For example, as illustrated in Figure~\ref{fig:parameter_level_example}, when changing the revolve angle, the chamfer feature on the top face is preserved because the chamfer operation explicitly references the face using the symbolic name \texttt{revolve\_1.top\_cap()}. As a result, modifications to upstream parameters do not necessarily require rewriting subsequent operations. In contrast, Python-based CAD representations such as CadQuery often rely on coordinate-based selections. When earlier operations are modified, the spatial configuration of the geometry may change, and downstream features that depend on explicit coordinates may need to be recomputed or redefined. Therefore, modifying earlier operations can invalidate downstream feature definitions or require substantial manual adjustments.


For LLM4CAD-DSL, we sample scripts from the LLM4CAD-DSL dataset. For CadQuery, we use the CAD-Coder dataset \cite{doris2025cadcoderopensourcevisionlanguagemodel}, which contains CadQuery scripts derived from the DeepCAD dataset. We randomly sample 3,000 scripts from each dataset for evaluation. It should be noted that the geometries in the two datasets are not identical. LLM4CAD-DSL contains a wider range of operations, including revolve, fillet, chamfer, and pocket features, while the CAD-Coder dataset mainly consists of sketches followed by extrusion operations.

To ensure a fair comparison, we select 20 general editing instructions that can be applied to arbitrary CAD models. The full list of editing instructions is provided in Appendix~\ref{app:sample_instructions}. For the LLM4CAD-DSL representation, we use the fine-tuned LLM4CAD-Editor model. Fine-tuning is necessary because LLM4CAD-DSL is a newly introduced domain-specific language that does not appear in the pretraining corpus of the base LLM. Without fine-tuning, the model would not have prior knowledge of the DSL syntax or entity selection mechanisms. For CadQuery, we directly use the pretrained Qwen3-32B model without additional fine-tuning. This choice reflects the fact that CadQuery is implemented in Python, which is widely represented in the LLM's pretraining data. As a result, the base model already possesses substantial prior knowledge of Python syntax and programming patterns, making it possible to generate CadQuery scripts without task-specific fine-tuning. For each script in the testing set, we randomly sample one editing instruction and ask the model to generate the edited script. We then measure the parsing rate of the generated scripts as an indicator of editing robustness.

Figure~\ref{fig:comparison_parsing_rate} presents the parsing rate comparison. LLM4CAD-DSL achieves a parsing rate of 82.5\%, significantly outperforming CadQuery, which achieves 58.4\%. This result suggests that the structured geometry entity selection mechanism in LLM4CAD-DSL makes the representation more robust for instruction-guided editing. In contrast, Python-based CAD scripts often rely on explicit 3D coordinate computations or tightly coupled procedural dependencies. When modifications are introduced, these dependencies can easily be broken, leading to invalid scripts or geometry generation failures. Overall, the results demonstrate that DSL-based CAD representations with explicit geometric references provide stronger support for reliable CAD editing with large language models.

\subsubsection{Comparison of Qualitative Results}
\begin{figure}[h]
    \centering
    \includegraphics[width=\columnwidth]{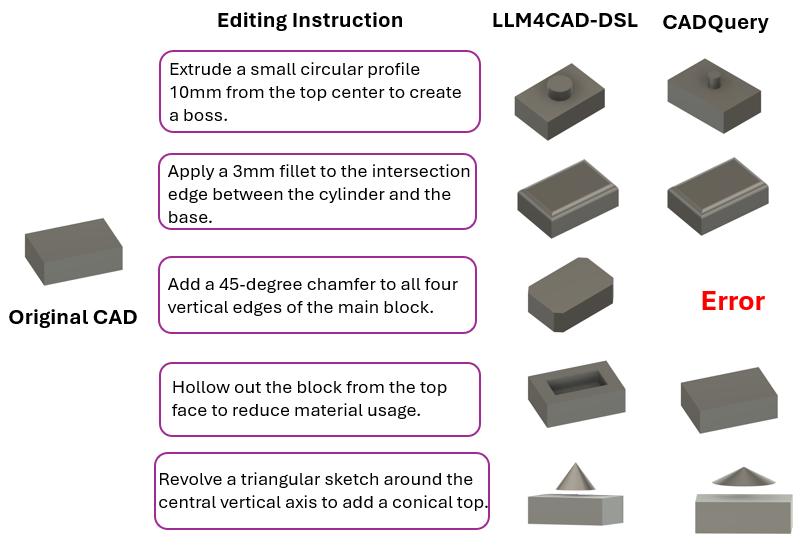}
    \caption{
Qualitative comparison between \textsc{LLM4CAD-DSL} and CadQuery 
}
    \label{fig:comparison_qualitative}
\end{figure}
Since the LLM4CAD-DSL and CAD-Coder datasets contain different geometries, direct side-by-side comparison is challenging. To enable a fair qualitative evaluation, we use simple block shapes as input to test both LLM4CAD-DSL and CadQuery. Figure~\ref{fig:comparison_qualitative} presents the qualitative evaluation results.
In the first two examples, both LLM4CAD-DSL and CadQuery successfully generate a cylindrical feature on the top surface and apply a fillet to the corresponding top edges. For simple geometric selection tasks in CadQuery, the LLM can rely on built-in selector strings such as  \texttt{\textgreater Z} to identify edges on the top surface.

However, when more complex selection is required, as illustrated in the third example, the limitations become evident. The LLM attempts to construct an auxiliary bounding box to isolate vertical edges, but the generated box is smaller than the actual block, resulting in an empty selection set and a subsequent runtime error. In contrast, within \textsc{LLM4CAD-DSL}, edges can be referenced directly through symbolic feature lineage (e.g., \texttt{extrude\_1(ridge(sketch\_1:vertex\_1))}), thereby avoiding the need for explicit geometric computations or spatial heuristics.

In the fourth example, the LLM using LLM4CAD-DSL successfully produces a hollow structure via a pocket operation. When generating CadQuery code, however, the LLM applies the \texttt{.shell(-thickness)} operation without correctly selecting the top face. As a result, all faces are implicitly considered, preventing the creation of a valid hollow region in the output CAD model.

In the fifth example, both LLM4CAD-DSL and CadQuery generate a cone positioned above the block, but the cone remains detached from the base geometry. This occurs because the triangular sketch is created on a newly defined vertical workplane rather than being anchored to an existing face of the solid. While the instruction specifies revolving a triangular sketch around the central vertical axis to form a conical top, it leaves the spatial reference for sketch placement underspecified. As a result, interpreting the instruction requires inferring an appropriate spatial reference in three-dimensional space. This example demonstrates that, although LLM generally follows the intended high-level design intent, precision in 3D geometric reasoning—such as spatial grounding and feature alignment—can be challenging when instructions do not explicitly define all spatial constraints. 

Overall, these qualitative results reveal a fundamental difference between symbolic and geometry-driven CAD generation paradigms. While CadQuery requires the LLM to reason about spatial relationships, topology traversal, and kernel-specific feature semantics, LLM4CAD-DSL provides structured abstractions that allow entities to be referenced through feature history rather than geometric inference. As task complexity increases, this distinction becomes more pronounced, with geometry-based code generation exhibiting higher sensitivity to selection ambiguity, incorrect workplane grounding, and feature mis-specification. These observations suggest that improving symbolic grounding and feature-level referencing is critical for scaling LLM-based CAD generation to more complex design scenarios.

\section{Limitations and Future Work}

Despite encouraging results, this study has several limitations.
First, the editing dataset is constructed through a data augmentation strategy that removes operations from existing programs to form source–target DSL pairs. By swapping the original and edited programs, the dataset implicitly includes feature addition tasks. However, this construction process does not adequately capture more complex editing scenarios that simultaneously require both addition and deletion of geometric features (e.g., replacing a cylindrical feature with a prismatic one). As a result, the current benchmark may underestimate the difficulty of realistic CAD editing workflows.

Second, our evaluation focuses primarily on one-step editing tasks. In practical design settings, CAD modeling is inherently iterative, involving long construction sequences with many variables, symbolic references, and dependent operations. As CAD programs grow longer, LLM-based editors may face a long-context dependency-tracking challenge: previously defined variables, feature names, and symbolic entity references must be retrieved and used consistently when later operations depend on earlier modeling steps. This concern is related to the findings of LoCoBench~\cite{qiu2025locobench}, which shows that long-context software engineering requires not only code generation, but also dependency traversal, cross-file reasoning, and multi-session memory retention across large codebases. In CAD editing, analogous failures may appear as forgotten feature references, inconsistent reuse of symbolic variables, or broken downstream operations, especially for complex designs such as metamaterial structures with repeated local patterns and long-range dependencies. However, our current evaluation does not examine the stability and performance of large language models under continuous editing-chain scenarios where instructions are provided incrementally over multiple steps.

Future work can address these limitations in several directions. One promising direction is to construct more realistic editing benchmarks that include composite editing instructions involving both feature addition and deletion, such as feature replacement, topology restructuring, and constraint-preserving modifications. Such tasks would better reflect practical CAD editing scenarios where design intent evolves through structural changes rather than isolated feature insertions.

Another important direction is to investigate iterative and long-context CAD editing, where a sequence of editing instructions is issued over time and each modification depends on the evolving geometric state of the model. Studying this setting would enable a more comprehensive evaluation of whether large language models can maintain stable performance, preserve design consistency, and correctly track feature dependencies, variables, and symbolic references across long editing trajectories.

Furthermore, future research may explore integrating geometry-aware verification mechanisms or feedback from CAD kernels to improve the robustness of generated editing programs. Such approaches could help detect infeasible operations, reduce topology-related failures, and better align generated edits with real-world design and manufacturability constraints.

\section{Conclusion}

This work presents LLM4CAD-Editor, an intent-aware framework for instruction-guided CAD editing that bridges high-level design intent and low-level parametric modeling operations. By leveraging the structured representation of LLM4CAD-DSL, the proposed approach enables large language models to perform precise and robust modifications through symbolic entity referencing and feature-level reasoning. 

To support training and evaluation, we construct a multimodal CAD editing dataset using DSL-based data augmentation and vision–language instruction synthesis. Experimental results show that the fine-tuned model achieves high parsing rates and strong geometric accuracy for parameter-level and operation-level editing tasks. For functional-level instructions, LLM-assisted evaluation demonstrates that the model can effectively infer appropriate geometric modifications to satisfy abstract design goals. Comparative experiments further reveal that DSL-based representations provide substantially improved editing robustness compared with Python-based CAD scripting approaches.

Overall, the findings suggest that reliable LLM-based CAD editing requires not only program synthesis capability but also structured representations that enable stable geometric grounding and intent-aware reasoning. The proposed framework provides a practical step toward more flexible and intelligent CAD systems that can support iterative design workflows and accommodate diverse user expertise levels. Future work will explore more realistic multi-step editing scenarios, richer composite editing operations, and geometry-aware verification mechanisms to further enhance the robustness and applicability of LLM-driven CAD editing.

\section*{Acknowledgments and Disclaimers}
This work was partially supported by the NVIDIA Academic Grant Program. We appreciate the computing resources, i.e., eight A100-80GB GPUs, provided by NVIDIA. 

The authors used ChatGPT (OpenAI, Instant 5.3), accessed via web browser, in March 2026 for language editing, paragraph revision and readability improvement only across multiple sections, including the Introduction, Discussion and Conclusion sections. The authors reviewed all edits and take full responsibility for the integrity and final wording of the manuscript.

\bibliographystyle{asmems4}  
\bibliography{asme2e} 

\appendix
\section*{Appendix}
\section{Qualitative Assessment of Editing Task}
\label{app:qualitative_assessment}

\subsection{Parameter-Level Results}
\begin{figure}[h]
    \centering
    \includegraphics[width=\columnwidth]{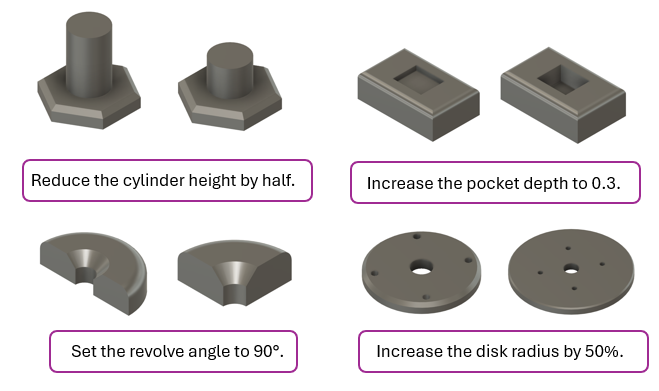}
    \caption{Examples of parameter-level editing tasks. In each subfigure, the left image shows the original CAD model, the right image displays the edited CAD model, and the text below provides the editing instruction.}
    \label{fig:parameter_level_example}
\end{figure}

Figure~\ref{fig:parameter_level_example} presents qualitative examples of parameter-level editing tasks. The subfigures illustrate several representative modifications, including changing the extrusion height, pocket depth, revolve angle, and circle radius in the sketch. These examples demonstrate that the fine-tuned editing model can accurately locate the relevant parameters in the DSL script and modify them correctly according to the user instruction.

This capability is further supported by the high parsing rate and IoU scores reported in Figure~\ref{fig:parsing_rate_results} and Figure~\ref{fig:IoU_results}. Together, these results indicate that the model can reliably handle parameter-level instructions while preserving the overall structure of the CAD program.

Parameter-level editing serves as a fundamental capability for CAD modification tasks, as it requires the model to precisely identify and update numerical values associated with geometric operations. The strong performance observed in these examples suggests that the model has successfully learned to perform localized parameter adjustments within the DSL scripts.

\subsection{Operation-Level Results}
\begin{figure}[h]
    \centering
    \includegraphics[width=\columnwidth]{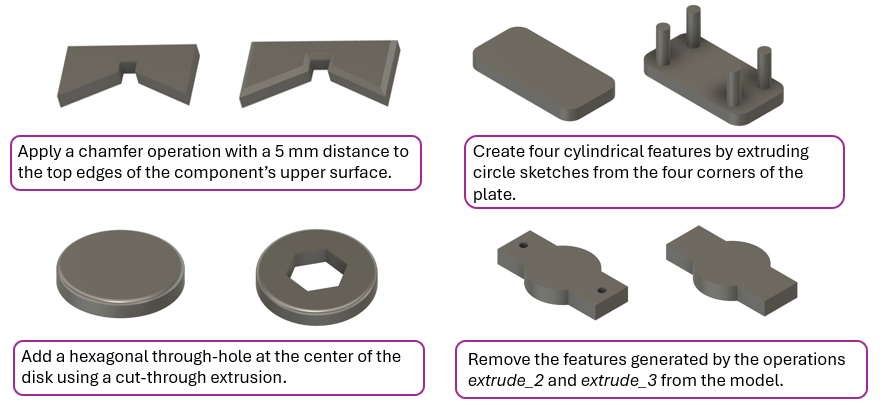}
    \caption{Examples of operation-level editing tasks.}
    \label{fig:operation_level_example}
\end{figure}

Figure~\ref{fig:operation_level_example} presents qualitative results for operation-level editing tasks. In operation-level instructions, the editing request typically specifies the type of geometric operation to be applied, such as chamfer, fillet, or extrusion, and may also include numerical parameters associated with the operation.

As shown in the examples, the model can successfully perform a variety of editing operations, including adding chamfers, fillets, holes, and prism features. These results demonstrate that the fine-tuned model is capable of identifying the relevant operations in the DSL script and generating the appropriate modifications to achieve the intended edit. Notably, the model can handle both addition-based edits and deletion-based edits, as illustrated in the last example of Figure~\ref{fig:operation_level_example}. This ability indicates that the model can flexibly modify the procedural structure of the CAD program rather than only adjusting existing parameters.

Operation-level instructions represent a middle level of design intent and are commonly used in real-world CAD editing workflows, where designers specify concrete geometric operations to modify the model. The qualitative results suggest that the model can effectively interpret and execute such instructions within the DSL editing framework.

\subsection{Functional-Level Results}

\begin{figure}[h]
    \centering
    \includegraphics[width=\columnwidth]{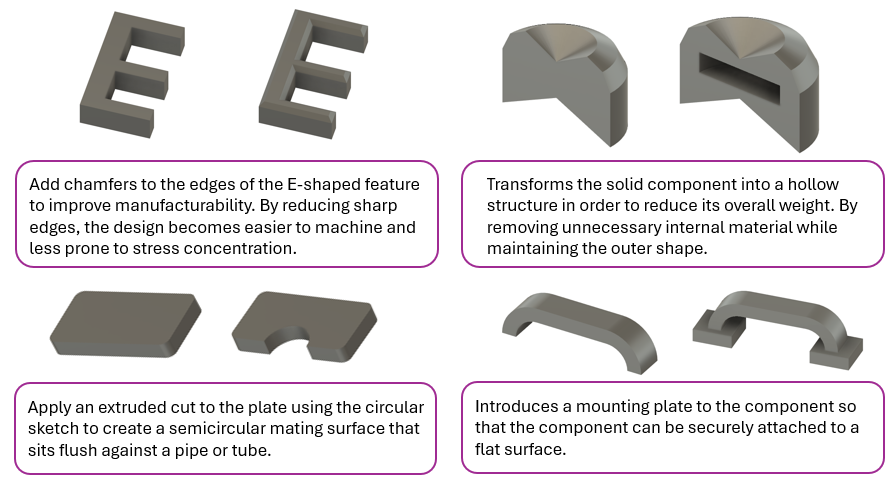}
    \caption{Examples of functional-level editing tasks.}
    \label{fig:functional_level_example}
\end{figure}

Figure~\ref{fig:functional_level_example} presents qualitative examples of functional-level editing tasks. As a high-level form of design intent, functional instructions describe the objective of the modification (e.g., “improve manufacturability”) rather than specifying explicit operations or numerical parameters.

For example, in the second case, the instruction aims to reduce the weight of the component by transforming the solid structure into a hollow one. The fine-tuned model successfully introduces a disk-shaped hollow feature inside the component while preserving the external geometry, demonstrating its ability to infer appropriate editing operations from abstract design goals.

After fine-tuning with high-level instructions, the model is able to interpret functional requests such as improving manufacturability, adding mounting interfaces, or introducing cut features at appropriate locations. This capability enables the system to translate high-level design intentions into concrete CAD modifications, which can be particularly helpful for non-expert users who may not be familiar with specific CAD operations or parameters.

\subsection{Lower-Quality Results}

\begin{figure}[h]
    \centering
    \includegraphics[width=\columnwidth]{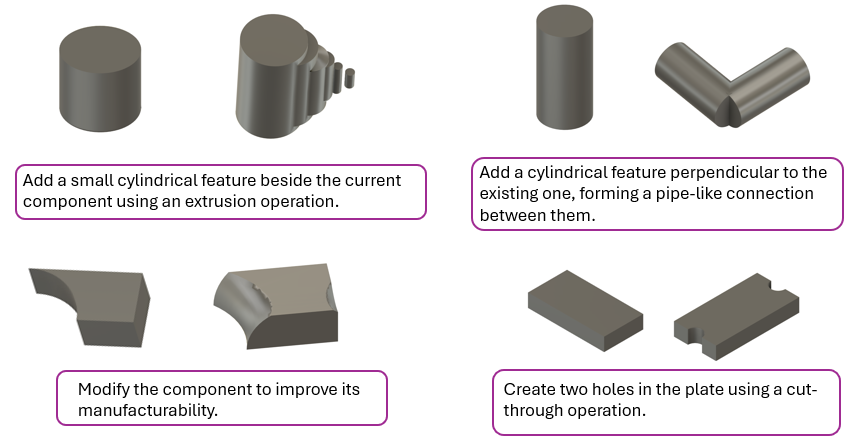}
    \caption{Examples of lower-quality cases.}
    \label{fig:low_quality_generation_example}
\end{figure}

Figure~\ref{fig:low_quality_generation_example} presents several examples of lower-quality editing results. These cases receive LLM-assisted ratings of 1 or 2, indicating that the generated edits fail to adequately satisfy the intended functional requirement. Through qualitative analysis, we observe two common sources of errors: over-editing and incorrect geometric reasoning.

The first issue is over-editing, where the model performs unnecessary or excessive modifications. For example, in the first case, when instructed to add a cylinder beside the existing one, the model generates several cylinders instead of a single feature. Similarly, in the third example, when asked to improve manufacturability, the model introduces a groove operation rather than using simpler operations such as chamfer or fillet. This modification results in a more complex surface geometry that does not effectively improve manufacturability.

The second issue is inaccurate geometric reasoning during feature construction. In the second example, when instructed to add a perpendicular cylinder to form a pipe-like connection, the model correctly generates a perpendicular cylinder and aligns the axes of the two features. However, the resulting geometry contains a break in the final CAD model because the model does not properly account for the cylinder radius during the connection. In the fourth example, when asked to add holes to the plate, the model places the hole center on the plate boundary, resulting in a semi-circular hole instead of a valid circular opening.

Overall, these lower-quality cases highlight the remaining challenges in instruction-guided CAD editing. While the model can often interpret high-level editing intents, it may struggle with precise geometric reasoning and constraint awareness, particularly when multiple geometric relationships must be satisfied simultaneously. Addressing these limitations may require improved geometric constraint modeling or additional training data that emphasizes spatial reasoning in CAD operations.

\section{General Editing Instructions for Comparison Experiment}
\label{app:sample_instructions}

The following are 20 example instructions from our editing dataset:

\begin{enumerate}
    \item Extrude a small circular profile 10mm from the top center to create a boss.
    \item Create a rectangular pocket on the front face with a depth of 5mm.
    \item Revolve a triangular sketch around the central vertical axis to add a conical top.
    \item Apply a 3mm fillet to the intersection edge between the cylinder and the base.
    \item Add a 45-degree chamfer to all four vertical edges of the main block.
    \item Extrude a circle through the entire body to create a clear passage.
    \item Use the revolve tool to cut a semi-circular groove around the outer surface.
    \item Create a hexagonal pocket on the side face that stops halfway through the part.
    \item Fillet all sharp external corners of the model to a radius of 2mm.
    \item Use a pocket operation with a draft angle to create a sloping interior cavity.
    \item Add a feature on the top surface that can serve as a mounting pillar for a PCB.
    \item Remove material from the center of the part to make it lighter while keeping the frame.
    \item Round off all sharp edges of the model so it is safe for a user to grip.
    \item Create an opening at each corner of the base to allow for M6 bolt installation.
    \item Add a smooth transition at the base joint to reduce stress concentration.
    \item Add a cylindrical support to the bottom to increase the part's height.
    \item Create a semi-circular groove on the side for a finger to rest in.
    \item Round the inner edge of the top hole to make it easier to insert a pin.
    \item Hollow out the block from the top face to reduce material usage.
    \item Create a flat, recessed area on the bottom so the part sits stable.
\end{enumerate}

\end{document}